# Carbon doped symmetric GaAs/AlGaAs quantum wells with hole mobilities beyond $10^6$ cm$^2$/Vs

*C. Gerl, S. Schmult, H.- P. Tranitz, C. Mitzkus, W. Wegscheider*

*Universität Regensburg, Institut für Experimentelle und Angewandte Physik,*

*93040 Regensburg*

**Abstract :**

Utilizing a novel carbon doping source, we prepared two-dimensional hole gases in a symmetric quantum well structure in the GaAs/AlGaAs heterosystem. Low temperature hole mobilities up to $1.2 \times 10^6$ cm$^2$/Vs at a density of $2.3 \times 10^{11}$ cm$^{-2}$ were achieved on GaAs (001) substrates. In contrast to electron systems, the hole mobility sensitively depends on variations of the quantum well width and the spacer thickness. In particular an increase of the quantum well width from an optimal value of 15 nm to 18 nm is accompanied by a 35 % reduction of the hole mobility. The quality of ultrahigh-mobility electron systems is not affected by the employed carbon doping source.

The almost perfect lattice match between Galliumarsenide (GaAs) and Aluminumarsenide (AlAs) allows for the preparation of heterostructures from these materials and their alloys without limitations with respect to layer thicknesses and composition. The concept of band-structure engineering can be therefore fully exploited. One prominent example are high electron mobility two-dimensional electron gases (2DEGs) exceeding mobility values of $10^7$ cm$^2$/Vs [1]. Such high electron mobilities have been proven to be essential for a variety of important discoveries including the fractional Quantum Hall effect [2], the stripe phases forming at high filling factors [3], or the recently reported excitonic condensation in double layer systems [4] and the microwave induced zero resistance states [5]. While the epitaxial growth of high mobility 2DEGs in the GaAs/AlGaAs material system is a state of the art technique today, the growth of holes systems with similar properties remains a fundamental challenge. Beryllium acts as an acceptor in (001) GaAs and AlGaAs, with the disadvantage of strong segregation employing typical growth conditions for high quality epitaxy [6]. Silicon serving as the standard donor in the considered material system for many growth directions, can also yield hole doping for growth on (311)A GaAs substrates [7]. Two-dimensional hole gases (2DHGs) prepared by this technique exhibit mobilities of up to 1.2 x $10^6$ cm$^2$/Vs [8]. However Si-doped GaAs (311)A 2DHGs show a pronounced mobility anisotropy which has been attributed to anisotropic interface roughness scattering rather than to the hole subband anisotropy. Furthermore, high mobility electron and hole systems cannot be combined in the same structure. Although carbon doping by means of CBr$_4$ and similar types of gas sources yield high hole concentrations up to $10^{20}$, cm$^{-3}$ such attempts using carbon containing precursors are incompatible with high mobility epitaxy. Reuter et al [9] realized p-doping with carbon evaporated from a graphite filament by electron beam bombardment. In this way, 2DHGs exhibiting mobilities up to 1.6 x $10^5$ cm$^2$/Vs were realized, in which a pronounced Rashba-induced spin splitting as well as clear signatures of fractional quantum Hall effect states were observed [10].

In this letter we report on C-doped 2DHGs, which have been grown in a dedicated high mobility MBE system capable of achieving electron mobilities of more than $10^7$ cm$^2$/Vs. A double sided modulation doped quantum well shows a low-temperature mobility of 1.2 x 10$^6$cm$^2$/Vs, which compares well with the so far reported record mobility on (311)A substrates [8]. Magnetotransport studies on our structures are characterized by a large number of integer and fractional QHE states. From a pronounced beating pattern in the low magnetic field Shubnikov-de-Haas oscillations, the occupation of the two Rashba-induced spin split heavy hole subbands was extracted. The carrier imbalance of these subbands was determined to 4 %, which is a strong indication for a highly symmetric quantum well. The dependence of the hole mobility on the quantum well width and on the carrier density measured in gate structures shows an unusual behavior and points towards the importance of intersubband scattering in these high mobility hole systems.

We have prepared a series of double sided symmetrically modulation doped quantum well structures with variations in quantum well width w and spacer thickness d on GaAs (001) substrates by using an electrically heated carbon sublimation source [11]. The design of this doping source allows a rapid heating of the graphite filament to the operation temperature in about 85 seconds. For a bulk carrier concentration of 9 x 10$^{18}$ cm$^{-3}$ in GaAs (growth rate 1 μm/hour), a heating power of 360 W, including parasitic power dissipation due to serial resistances, is required. A detailed schematic of the sample layer sequence can be found in Fig. 1. The 2DHG is located d + 100 nm below the surface. All samples have been grown using the optimized growth temperature for high mobility 2DEGs of 640 °C.

Transport measurements have been carried out in van der Pauw and Hall bar geometry either in a $^4$He cryostat at a temperature of T = 1.5 K or in a $^3$He/$^4$He dilution refrigerator cryostat at a bath temperature below 30 mK. Magnetoresistance measurements were performed by means of standard lock-in techniques with operating currents <100 nA. On one sample a homogeneous aluminum top-gate was evaporated.

Figure 2 shows the dependence of the mobility and the density of the 2DHGs on the quantum well width and the spacer layer for identical doping concentrations. Similar to 2DEG systems a decrease of the carrier density with an increasing thickness of the spacer layer is seen. While the hole density does not show a significant dependence on the well width (Fig 2a), a clear peak arises when plotting the mobility as a function of this quantity. The decreasing mobility for wide quantum wells, which is markedly different compared to high mobility 2DEGs, implies that intersubband scattering has to be taken into account for these high mobility 2DHGs, since the subband energy difference between confined light and heavy hole states is small for wide wells. In contrast, scattering due to both layer interfaces seems to limit the mobility in narrow quantum wells. This has led to an optimized geometry with a well width w of 15 nm and a spacer thickness d of 80 nm exhibiting a 1.5 K mobility of $9.1 \times 10^5$ cm$^2$/Vs at a density of $2.3 \times 10^{11}$ cm$^{-2}$ (Sample A).

In Fig. 3 the longitudinal and transversal magnetoresistance of sample A processed in a Hall bar geometry are plotted for a measurement temperature of 30 mK. The integer Quantum Hall effect (QHE) as well as the fractional QHE are clearly recovered, manifested in the observation of a large number of fractional states i.e. 2/3, 3/5, 4/7 and 5/9. The pronounced minima at ν= 4/3 and 5/3 and the distinct features at 5/2 and 7/2 as well as the highly symmetric behavior of the Shubnikov-de-Haas (SdH) oscillations (inset of Fig. 3) around zero magnetic field are pointing out the high quality of the sample. The low temperature mobility of this sample has been determined to $1.2 \times 10^6$ cm$^2$/Vs. This represents an increase of 30 % in mobility with respect to the 1.5 K measurement.

Fourier analysis of the low-field (0.05<B<0.5) SdH data determines the occupation of the two subbands which can be assigned to heavy hole spin up and spin down subbands with densities $p_1 = 1.1 \times 10^{11}$ cm$^{-2}$ and $p_2 = 1.2 \times 10^{11}$ cm$^{-2}$ respectively. The carrier imbalance $\Delta p = (p_1 - p_2)/p_t$ with $p_t = 2.3 \times 10^{11}$ cm$^{-2}$ is only 4 % which indicates a rather small Rashba coefficient for this structure [12]. Since the Rashba-induced spin splitting is due to structural inversion asymmetry, this demonstrates that a highly symmetric quantum well has been achieved, indeed. From the two different doping densities of the upper and the lower doping layer we can estimate, that about 30 % of the upper doping layer contribute to the hole carrier density in the quantum well.

Measurements in Hall bar geometry of a device equipped with a top gate (Sample B) can be seen in Figure 4. With a gate potential of zero volts relative to the 2DHG, a density reduction of $0.8 \times 10^{11}$ cm$^{-2}$ is observed compared to samples without gate. We attribute this fact to an elevated Fermi level across the sample due to band pinning with evaporated gate. Applying a positive voltage to the gate with respect to the quantum well leads to a depletion of the charge carriers down to $1.2 \times 10^{11}$ cm$^{-2}$. No further depletion can be achieved for higher gate voltages. This gives us a quantity for the contribution of the delta doping below the 2DHG to the total hole density of $p_t = 2.3 \times 10^{11}$ cm$^{-2}$ and leads to almost identical densities resulting from the upper and the lower modulation doping. This confirms the high symmetry of the quantum well and, thus, the absence of a large Rashba-induced spin splitting. Applying a negative potential to the gate increases the density as expected for hole systems. The mobility rises with increasing density up to $9.0 \times 10^5$ cm$^2$/Vs at $2.1 \times 10^{11}$ cm$^{-2}$. However for even higher densities the mobility drops significantly. This unusual effect is in contradiction to the typically observed increase of the mobility with increasing density as a result of improved screening. We account this effect to the reduced energetic difference between the Fermi energy and the light hole subband which could result into a more efficient scattering channel from the heavy hole states into the light hole states.

In conclusion, we demonstrated high mobility two-dimensional hole gases in the GaAs/AlGaAs heterosructure system with a mobility of $1.2 \times 10^6$ cm$^2$/Vs at a density of $2.3 \times 10^{11}$ cm$^{-2}$. A novel carbon doping source has been used for the fabrication of double sided modulation doped quantum well structures without apparent segregation of carbon. In addition, no effect on the mobility of electron gases has been seen to date. Due to the high mobilities of these 2DHGs, effects like a pronounced fractional QHE and the contribution of structure inversion asymmetry, so far mainly limited to the (311)A surface, can now be observed on (001) faced GaAs.

The work was financially supported by Deutsche Forschungs Gemeinschaft (GK 638) and BMBF via the program framework "Spinelektronik und Spinoptoelektronik in Halbleitern".

**References:**

1   M. A. Zudov, R. R. Du, L. N. Pfeiffer, K. W. West, Phys. Rev. Lett. **90**, 046807 (2003).

2   J. P. Eisenstein and H.L. Stormer, Science **248**, 1510 (1990).

3   J. P. Eisenstein, K. B. Cooper, L. N. Pfeiffer, K. W. West, Phys. Rev. Lett. **88**, 076801 (2002).

4   M. Kellogg, J. P. Eisenstein, L. N. Pfeiffer, K. W. West, Phys. Rev. Lett. **93**, 036801 (2004).

5   R. G. Mani, J. H. Smet, K. v. Klitzing, V. Narayanamurti, W. B. Johnson, V. Umansky, Nature **420**, 646, (2002).

6   S. V. Ivanov, P. S. Kop'ev, N. N. Ledentsov, J. Cryst. Growth **108**(3-4), 661 (1991).

7   M. Henini, P. J. Rodgers, P.A. Crump, B.L. Gallagher, G. Hill, J. Cryst. Growth **150** (1-4), 446 (1995).

8   J. J. Heremans, M. B. Santos, M. Shayegan, Appl. Phys. Lett. **61**(14), 1652 (1992).

9   D. Reuter, A.D. Wieck, Rev. Scient. Instr. **70**, 8, 3435 (1999).

10  B. Grbic, C. Ellenberger, T. Ihn, and K. Ensslin, D. Reuter and A. D. Wieck, Appl. Phys. Lett. **85**, 12 (2004).

11  Modified SUKO40 carbon source fabricated by Dr. Eberl MBE-Komponenten GmbH, Gutenbergstrasse 8, D-71263 Weil der Stadt, Germany, (http://www.mbe-kompo.de).

12  S. J. Papadakis, E. P. De Poortere, H. C. Manoharan, M. Shayegan, R. Winkler, Science **283**, 2056 (1999).

**Figure captions:**

Fig. 1: Schematic structure of double sided modulation doped quantum well sample. After a GaAs buffer layer and a superlattice, a GaAs quantum well is grown, sandwiched between two $Al_xGa_{1-x}As$ (x = 0.33) layers including a C-$\delta$-doping step. A final GaAs layer of 10 nm serves as a cap. The sketch is not to scale.

Fig. 2: Hole mobility and density as a function of quantum well width w (a) and spacer thickness d (b) measured in van der Pauw geometry at 1.5 K. Sample A and B are in focus of the displayed measurements. The lines are a guide to the eye.

Fig. 3: Longitudinal and Hall resistance at a bath temperature <30 mK (Sample A). The Hall bar has a length of
1 mm and a width of 200 µm. The mobility is $1.2 \times 10^6 cm^2/Vs$ at a density of $2.3 \times 10^{11} cm^{-2}$. Inset: Low magnetic field (-0.5 T < B < 1 T) Shubnikov-de-Haas oscillations and according Hall measurement.

Fig. 4: Mobility and gate voltage as a function of hole density measured at 1.5K. Leakage current is below 50 pA (Sample B). The lines are a guide to the eye.

Fig.1:  (C. Gerl et al.)

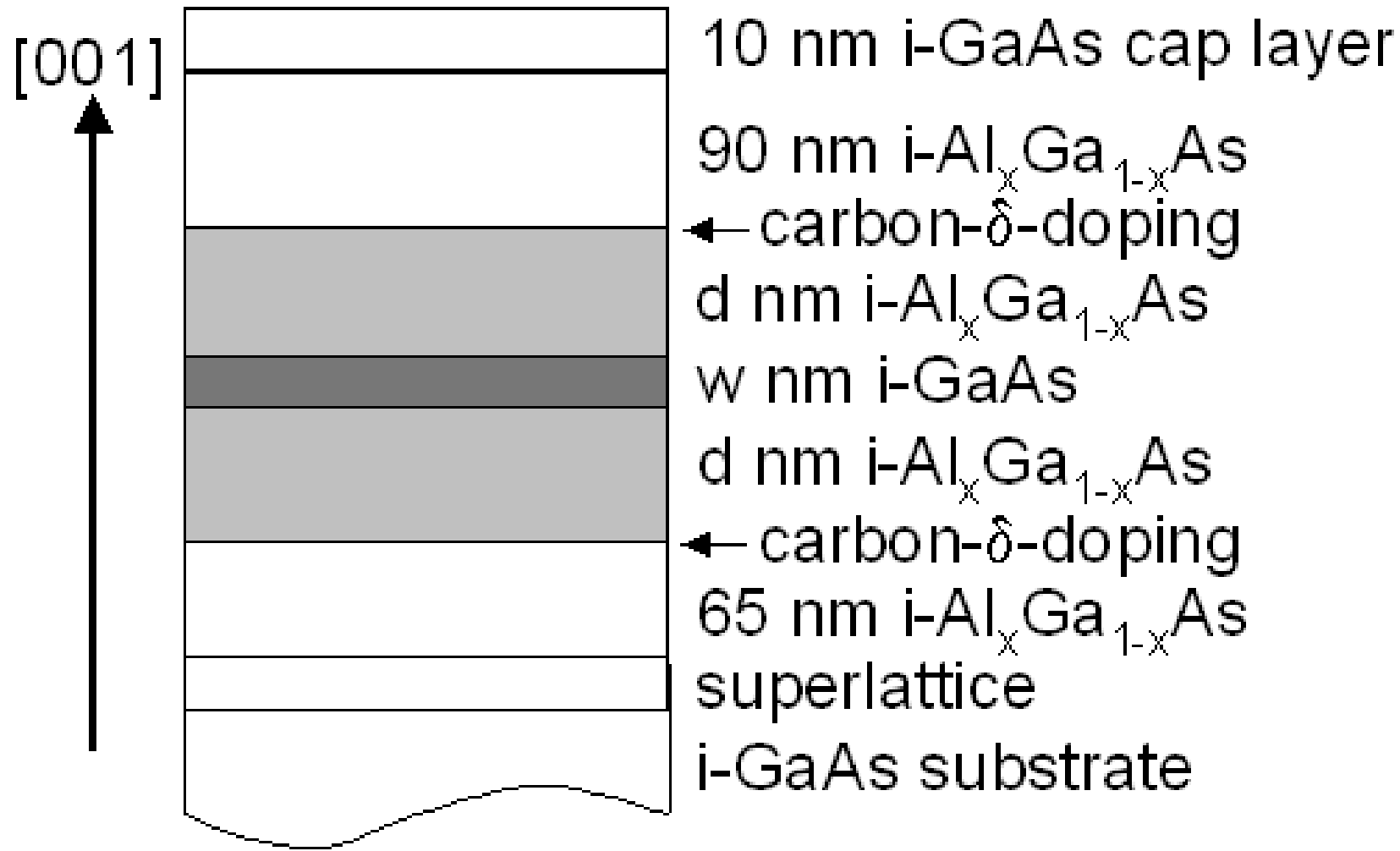

**Fig. 2:** (C. Gerl et al.)

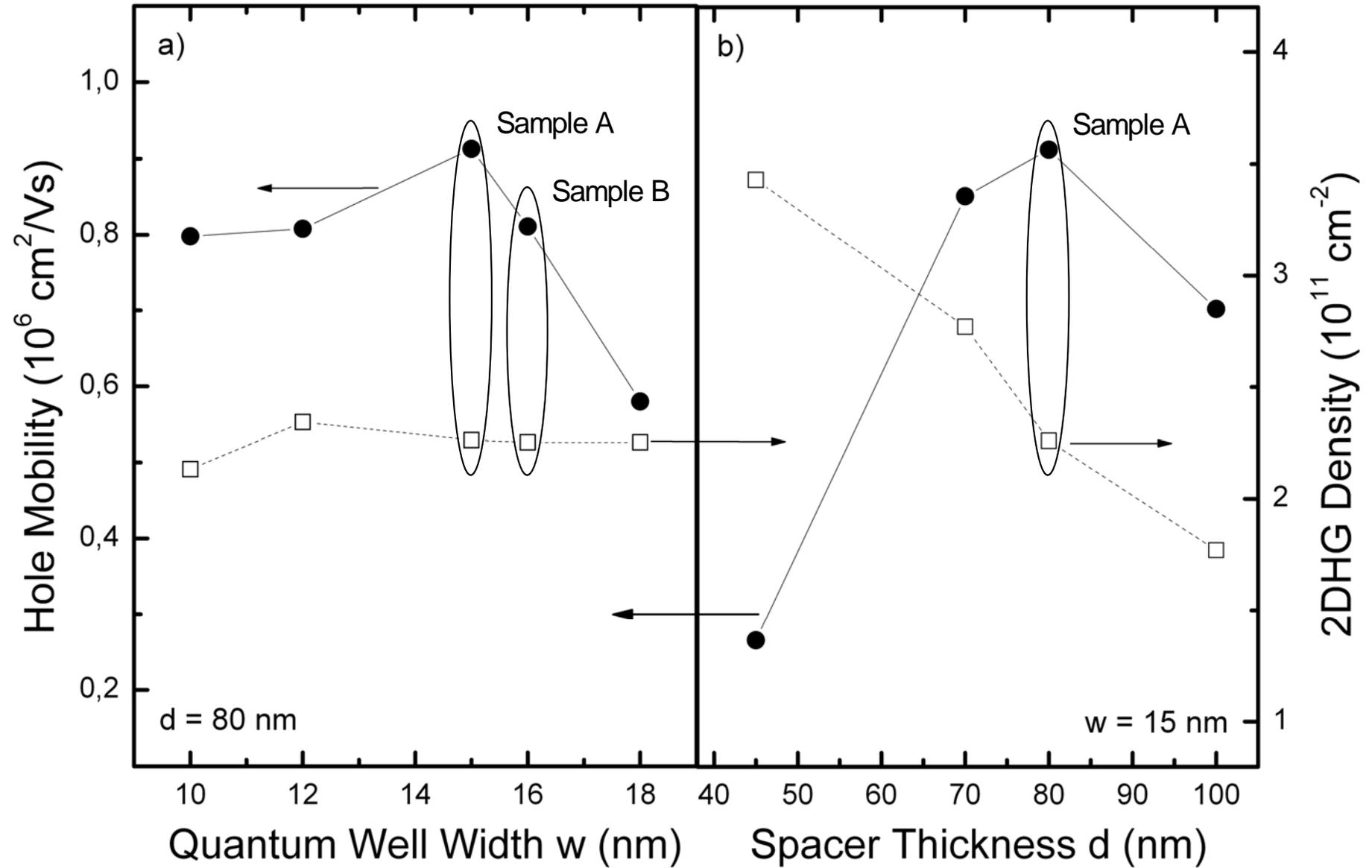

**Fig. 3:** (C. Gerl et al.)

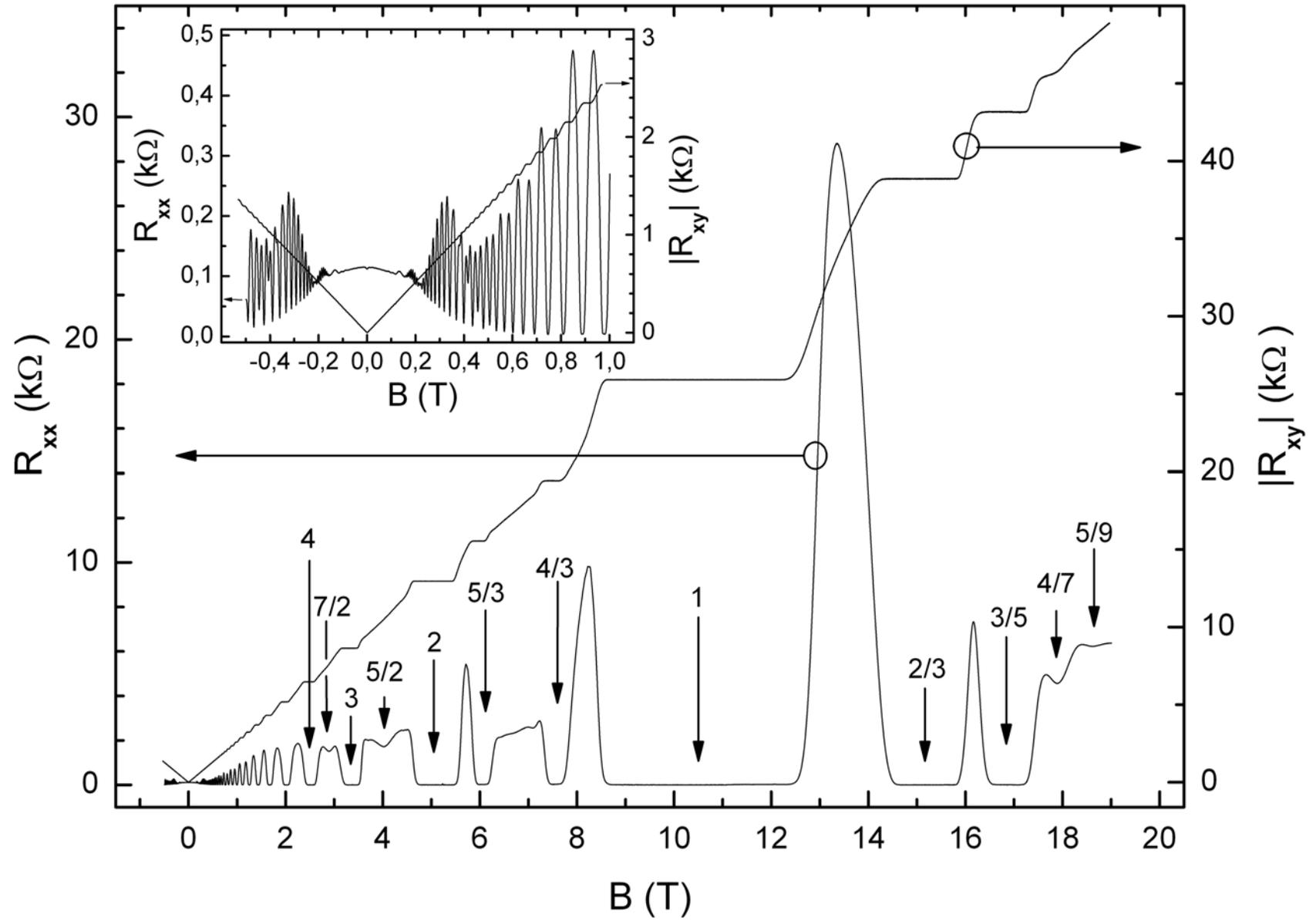

**Fig. 4:** (C. Gerl et al.)

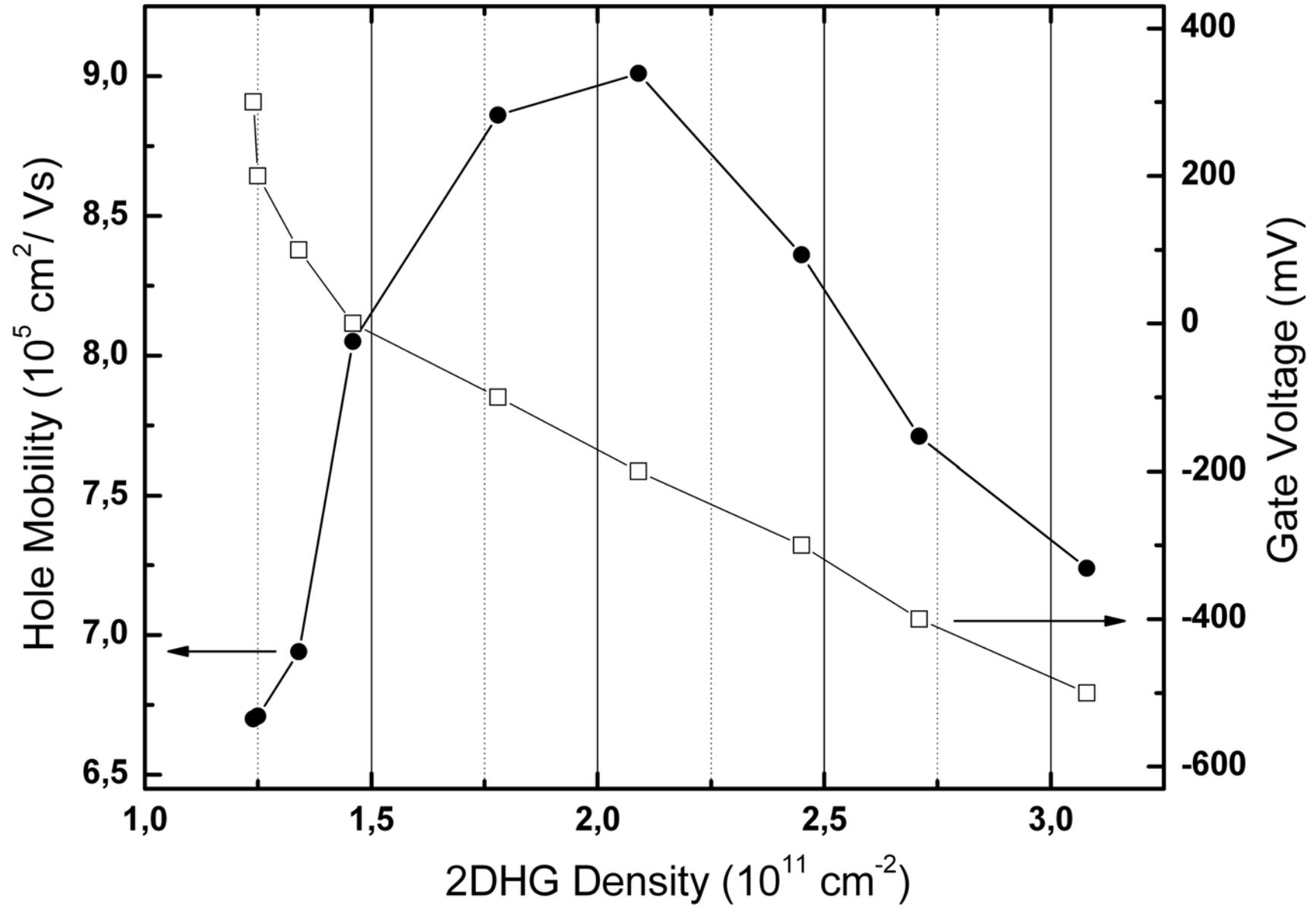